\documentclass[preprint,12pt,a4paper,showpacs]{revtex4}
\usepackage{amssymb}
\usepackage{latexsym}

\newcommand{\dd}{\textrm{d}}
\newcommand{\const}{\textrm{const}}

\begin{document}

\title{Singular disk of matter in the Cooperstock--Tieu galaxy model}
\author{Miko\l{}aj Korzy\'nski}
\affiliation{Institute of Theoretical Physcis, Warsaw University, ul. Ho\.za 69, 00-681 Warsaw, Poland}
\email{mkorz@fuw.edu.pl}

\begin{abstract}

Recently a new model of galactic gravitational field, based
on ordinary General Relativity, has been proposed 
by Cooperstock and Tieu  in which no 
exotic dark matter is needed to fit the observed rotation curve
to a reasonable ordinary matter distribution.  
We argue that in this model the gravitational field is generated 
not only by the galaxy matter, but by a thin, singular disk as well.
The model should therefore be considered unphysical.

\end{abstract}


\maketitle

\section{Introduction}

In \cite{ct} the authors derived a set of non-linear
equations which describes a
gravitationally bound rotating cloud of dust in General Relativity.
Basing on their result, the authors  propose a model of galaxy to which
a flat rotation curve can be fitted without dark, unobservable matter. The result is then applied to several galaxies, including the Milky Way.

We shall argue that
\begin{enumerate}
\item the set of equations they derive is essentialy correct, though does not
posses  non--trivial, asymptotically flat solutions,
\item the model of galaxy they propose is questionable because it posseses an additional source
of gravitational field in the form of a rotating flat disk at $z=0$.  
\end{enumerate}

\section{Gravitationaly bound cloud of dust in GR}

In GR a gravitationally bound cloud of dust is described by 
the stress-energy tensor $T^{\mu\nu} = \rho\, u^\mu u^\nu$ such that
$u^\mu$ is the dust four--velocity vector and $\rho$ its density 
(see for example \cite{schutz}).
The metric satisfies the Einstein equations
\begin{equation}
G^{\mu\nu}[g_{\alpha\beta}] = 8\pi G\, \rho \,u^\mu u^\nu \label{einst}
\end{equation}
where the Einstein tensor field is to be considered a functional of the metric
field $g_{\alpha\beta}$ (we set $c=1$ throughout the paper).
Note that the Biachi identities immediately  imply that the flow
of the dust is geodesic and source--free
\begin{eqnarray}
\rho \,u^\mu \,\nabla_\mu u^\nu &=& 0 \\
\nabla_\mu(\rho u^\mu) &=& 0.
\end{eqnarray}  

Let $\eta$ denote the flat Minkowski metric. 
The conventional way to simplify this highly non--linear set of equations
is to perform a perturbative expantion of some of the quantities in
$G$:
\begin{eqnarray}
g_{\alpha\beta} &=& \eta_{\alpha\beta} + G\,g^{(1)}_{\alpha\beta} + G^2
\,g^{(2)}_{\alpha\beta} + O(G^3)\\
u^\alpha &=& u_{(0)}^\alpha + G\, u_{(1)}^\alpha + G^2\, u_{(2)}^\alpha + O(G^3)
\end{eqnarray}
and obtain the following equation
\begin{eqnarray*}
&&G\cdot G'^{\mu\nu}[g^{(1)}] + G^2\cdot\left(
G'^{\mu\nu}[g^{(2)}] + \frac{1}{2} G''^{\mu\nu}[g^{(1)},g^{(1)}]\right) + O(G^3) = \\
&=&8\pi G \,\rho (u_{(0)}^\mu + G\,u_{(1)}^\mu)(u_{(0)}^\nu + G\,u_{(1)}^\nu) + O(G^3)
\end{eqnarray*}
($G'^{\mu\nu}[\,\cdot\,]$ and $G''^{\mu\nu}[\,\cdot\, ,\,\cdot\,]$ denotes, respectively, the first and second 
functional derivative of $G^{\mu\nu}[g_{\alpha\beta}]$ for $g_{\alpha\beta} = \eta_{\alpha\beta}$).
We can now attempt to solve these equations order by order.

This approach, however, fails to reproduce the set of equations from \cite{ct}.
This is due to the fact that the exact solution families approximated there
depend on $G$ via $\sqrt{G}$, \emph{i. e.}~their dependence on the
coupling constant is non--analytic.
 Since we want to follow the approach of \cite{ct}, we shall 
expand (\ref{einst}) in $\sqrt{G}$ instead of $G$:

\begin{eqnarray}
g_{\alpha\beta} &=& \eta_{\alpha\beta} + G^{1/2}\,g^{(1)}_{\alpha\beta} + G
\,g^{(2)}_{\alpha\beta} + O(G^{3/2}) \\
u^\alpha &=& u_{(0)}^\alpha + G^{1/2}\, u_{(1)}^\alpha + G\, u_{(2)}^\alpha + 
O(G^{3/2})
\end{eqnarray}
and therefore
\begin{eqnarray*}
&&G^{1/2}\cdot G'^{\mu\nu}[g^{(1)}] + G\left(\,G'^{\mu\nu}[g^{(2)}]+
\frac{1}{2}G''^{\mu\nu}
[g^{(1)},g^{(1)}]\right) + O(G^{3/2})\\
 &=&8\pi G\,\rho \,u_{(0)}^\mu u_{(0)}^\nu + O(G^{3/2}) .  
\end{eqnarray*}
Rewritten order by order this is equivalent to
\begin{eqnarray}
G'\,^{\mu\nu}[g^{(1)}] &=& 0 \label{pert1}\\
G'\,^{\mu\nu}[g^{(2)}] + \frac{1}{2} G''\,^{\mu\nu}[g^{(1)},g^{(1)}] &=&
8 \pi \rho u^{(0)} u^{(0)} \label{pert2}.
\end{eqnarray}

The first equation is the linearized vacuum Einstein equation. 
If we impose additionally the Lorentz--De Donder gauge condition (see 
\cite{schutz}), (\ref{pert1}) is equivalent to 
\begin{equation} 
\square\, g_{(1)}^{\mu\nu} = 0
\end{equation}
($\square$ being the flat metric $\eta$ d'Alembert operator). In particular, for stationary problems
\begin{equation}
\Delta\,g_{(1)}^{\mu\nu} = 0 \label{laplace}.
\end{equation}


Note that for an asymptotically flat metric $g_{(1)}^{\mu\nu} \to 0$ for
$r \to \infty$, which together with (\ref{laplace}) implies that $g^{(1)} = 0$.
In other words, no asymptotically flat solutions are possible with
the metric depending on $\sqrt{G}$ in the lowest order.

\section{Cooperstock and Tieu galactic rotation curve models}

Cooperstock and Tieu assumed the metric to be of the form of
\begin{equation}
g = -e^{\nu}(dz^2 + dr^2) - r^2\,d\phi^2 +(dt - N d\phi)^2 \label{gg}
\end{equation}
with 
\begin{eqnarray}
N &=& r \frac{\partial \Phi}{\partial r} \label{n} \\
\Phi &=& e^{-k|z|}\,J_0(kr), \label{n+1}
\end{eqnarray}
both $\Phi$ and $N$ being of the order of $\sqrt{G}$. 
It is straightforward to verify that this metric fails to satisfy (\ref{pert1}),
as $\Delta g^{(1)}$ is not equal to zero, but has a distributional source 
proportional to $\delta(z)$. (\ref{gg}) is not, therefore, a well--defined, \emph{global} metric with rotating dust as the only source of gravity. 
 
In fact, it does not satisfy the first of the two equations derived by Cooperstock and Tieu:
\begin{eqnarray}
N_{rr} + N_{zz} - \frac{N_r}{r} &=& 0 \\
\frac{N_r ^2 + N_z ^2}{r^2} &=& 8\pi G\,\rho
\end{eqnarray}
As before, due to the presence of $|z|$ in (\ref{n+1}) the first equation has a distributional singularity at $z=0$ rather than 0 on
the right--hand side. In the following section we shall prove that this results in 
presence of an additional, distributional term in the stress--energy tensor $T^{\mu\nu}$. 

\section{Komar integral}

The shortest way to prove that the galaxy model in \cite{ct} has a sheet of matter at $z=0$ is to 
consider the Komar integral of the timelike Killing vector $\frac{\partial}{\partial t}$.
Define $\tau = g(\frac{\partial}{\partial t},\cdot)$. 
It is easy to prove in general that if $\nabla_\mu \tau_\nu + \nabla_\nu \tau_\mu = 0$, the following 
identity holds:
\begin{equation}
\dd \star \dd \tau = \frac{1}{3}\,R^{\mu\alpha}\,\tau_\alpha\,\epsilon_{\mu\nu\rho\sigma}\, \dd x^\nu\wedge
\dd x^\rho \wedge \dd x^\sigma
\end{equation}
($\epsilon$ is the volume 4--form; if the metric is vacuum the identity reduces to the Komar current conservation formula). We can
integrate it over an arbitrary 3--volume V: 
\begin{equation}
\int_{\partial V} \star\dd\tau = \int_V \dd\star\dd\tau  = \frac{4\pi G}{3} \int_V
(2 T^{\alpha\mu} \tau_\alpha - T\, \tau^\mu) \epsilon_{\mu\nu\rho\sigma}\,\dd x^\nu\wedge\dd x^\rho
\wedge\dd x^\sigma \label{cyl} .
\end{equation}
Take $V$ to be the cylinder defined by the following conditions: $r \leq R$, $-a \leq z \leq a$, $t = \const$, for given positive constants $a$ and $R$. 
The boundary of this volume consists of a side 2--surface $r=R$ and two circle--shaped 2--surfaces 
$z=a$ and $z=-a$. Let us take the limit of $a \to 0$, \emph{i. e.}
the flat cylinder limit. If $T^{\mu\nu}$ had no singularities but consisted solely of 
rotating dust, this limit would be zero, since the volume of the cylinder goes to zero as
we shrink it. It is, however, not the case in the metric (\ref{gg}). Indeed, the 
side surface integral of (\ref{cyl}) does converge to $0$, but the top and bottom side integrals $I_t$
and $I_b$ do not:
\begin{equation}
I_t + I_b = \int_0^{2\pi} \dd\varphi \int_{0}^R \dd r \,  \left.\frac{N}{r}\frac{\partial N}{\partial z}\right|_{z=a} - \int_0^{2\pi} \dd\varphi \int_{0}^R \dd r \,  \left.\frac{N}{r}\frac{\partial N}{\partial z}\right|_{z=-a} .
\end{equation}
With $N$ given by (\ref{n}) and (\ref{n+1}) $I_t$ and $I_b$ are in fact equal and therefore
\begin{equation}
I_t + I_b =  
 4\pi \int_0^R \dd r \, \left.\frac{N}{r}\frac{\partial N}{\partial z}\right|_{z=a} \to
 -4\pi k^3 \int_0^R \dd r \, r \,J_0'(kr)^2 \neq 0 .  
\end{equation}
The reason is, once again, that $\frac{\partial N}{\partial z}$ has a discontinuity at $z=0$ due to the presence
of $|z|$ in its definition. 

Note that the same argument can be applied to the axial Killing
vector $\frac{\partial}{\partial\varphi}$ as well. This indicates that the disk has both mass and
angular momentum, simmilarily to the singularity in the exact solution of van Stockum \cite{bonnor}. 

The solution should therefore be interpreted as possesing an additional, distributional source
of gravity at $z=0$ and the resulting discontinuity of the metric's first derivative as a physical
one rather than a coordinate system artifact, as is claimed in \cite{ct}.

\end{document}